\documentclass[12pt,a4paper]{spie}  %>>> use for US letter paper
\usepackage{amsmath,amsfonts,amssymb}
\usepackage{graphicx}
\usepackage[colorlinks=true, allcolors=blue]{hyperref}

\title{Using modern motion estimation algorithms\\ in existing video codecs }

\author{Daniel J. Ringis}
\author{Davinder Singh}
\author{Francois Pitie}
\author{Anil Kokaram}
\affil{Sigmedia Group, Electronic and Electrical Engineering Dept., Trinity College Dublin, Ireland}

\authorinfo{For further information please contact A. Kokaram :  anil.kokaram@tcd.ie}

\begin{document} 
\maketitle

\begin{abstract}
Motion estimation is a key component of any modern video codec. Our understanding of motion and the estimation of motion from video has come a very long way since 2000. More than 135 different algorithms have been recently reviewed by Scharstein et al 
http://vision.middlebury.edu/flow/. These new algorithms differ markedly from Block Matching which has been the mainstay of video compression for some time. This paper presents comparisons of H.264 and MP4 compression using different motion estimation methods. In so doing we present as well methods for adapting pre-computed motion fields for use within a codec. We do not observe significant gains to be had with the methods chosen w.r.t. Rate Distortion tradeoffs but the results reflect a significantly more complex interrelationship between motion and compression than would be expected. There remains much more to be done to improve the coverage of this comparison to the emerging standards but these initial results show that there is value in these explorations.
\end{abstract}

\keywords{Video Compression, Motion Estimation, Optic Flow, H.264, MPEG, MPEG4, Open Alliance}

\section{INTRODUCTION}
\label{sec:intro}  % \label{} allows reference to this section

Motion estimation is a key component of any modern video codec. Our understanding of motion and the estimation of motion from video has come a very long way since 2000. A quick look at the current evaluation of optic flow techniques first presented in 2011 by Baker, Scharstein, Lewis et al\cite{Baker_2011} and the corresponding online league table\footnote{{\tt http://vision.middlebury.edu/flow/}} show a comparison of 135 different algorithms. For frame interpolation accuracy, block based algorithms are at rank 41 at best. In the Sintel dataset for optic flow comparison\cite{Butler:ECCV:2012} they show up as rank 137 out of 148 different algorithms. While this does not mean that better motion estimation techniques combined with rate control within a codec would have real impact, we believe its worth the experiment. Our contribution to the existing body of literature in motion estimation for encoding is to inform the ongoing debate of the impact of motion estimation performance on encoding performance, a debate ongoing since the 1990's.

%https://pdfs.semanticscholar.org/presentation/95c9/8270f6d2da06262808909b7b5fd088cef7bc.pdf

\subsection{Motion Estimation in Video Compression}
The heart of all hybrid video compression schemes is a method for predicting pixels from spatial or temporal locations in the current or other frames. The idea of exploiting temporal redundancy for prediction dates back to early experiments by Haskell et al in the 1970's\cite{haskell_1971, haskell_1976}. The image sequence model used for prediction has remained unchanged since then as follows.
\begin{equation}
    I_n({\bf x}) = I_{n-1}({\bf x} + {\bf d}_{n,n-1}({\bf x})) + e_n({\bf x}) \label{mpred}
\end{equation}
In this model the pixel value at site ${\bf x}$ in frame $n$, $I_n({\bf x})$,  is predicted by the value of the pixel at location ${\bf x} + {\bf d}_{n,n-1}({\bf x})$ in a previous frame $n-1$, $I_{n-1}({\bf x} + {\bf d}_{n,n-1}({\bf x}))$ allowing for some prediction error $e({\bf x})$ at that site ${\bf x}$ in frame $n$. That error or residual is also called the motion compensated residual. We say that $I_{n-1}$ is motion compensated with respect to frame $n$ when it is shifted according to the motion between the frames ${\bf d}_{n,n-1}({\bf x})$ at the relevant site ${\bf x}$.
Data compression is achieved by transform coding of the residual $e_n(\cdot)$.  

This model is well known to fail in occluded/uncovered areas and so the prediction equation can be extended to support from the next as well as previous frames. This bi-directional prediction mode was introduced quite early on in the development of video codecs and incorporated into MPEG2,4 H.264. The current evolution of compression standards with VP9, H.265, H.266 and AV1 have extended the idea further to allow prediction from any other kind of supporting frame, be it in the sequence or auxiliary to it.

To generate a solution for the motion vector ${\bf d}_{n,n-1}$ in video compression the key assumption has been that the motion field is piecewise constant. Hence motion is constant over a block of a size that varies from $2 \times 2$ pixels up to $128 \times 128$ pixels in recent standards. The codec adaptively chooses both that block size and the motion vector associated with that block.  This class of techniques is known as Block Matching. Efficient search algorithms have been explored since 1979 when gradient based search was first introduced by Netravali and Robbins\cite{netravali_1979}.  Since then, the estimation of ${\bf d}_{n,n-1}$ has been dominated by direct search techniques for solving equation~\ref{mpred} e.g. Logarithmic search\cite{jain_1981}, Diamond Search \cite{tham_1998}, Sucessive Elimination\cite{sea_1995}, 3DBM\cite{DEHAAN1994229} and Hexagon Search Pattern\cite{zhu_2002}. Many of these are summarised here\cite{liyin_2010, sinha_2003}.

The optimisation cost function in block matching for choosing the best motion vector has traditionally been some variant of the sum of the absolute or squared residuals/prediction error over a block, SAE or SSE. For video compression, Girod\cite{girod_1994} first presented rate distortion criteria appropriate for motion estimation in a codec. That has since become part of the optimisation scheme used in all codecs today\cite{sullivan_1998}. The essential idea is to choose a motion vector which minimises the prediction energy as well as the information needed to encode the motion vector itself. This motion estimation rate distortion energy $E({\bf d})$ is stated as follows.
\begin{equation}
    E({\bf d}) = \lambda R({\bf d}) + \sum_{{\bf x}\in B} |e_n({\bf x}, {\bf d})| \label{rd}
\end{equation}
where $B$ indicates a block, $|e_n({\bf x}, {\bf d})|$ is the SAE (for instance), $R({\bf d})$ is the rate or bits required to encode the vector ${\bf d}$ and $\lambda$ is a hyperparameter which controls the importance of rate control versus distortion management.

\subsection{Historical impact}
In 1979  Netravali and Robbins \cite{netravali_1979} attributed about 20\%-30\% improvement in bitrate and about 0.5dB improvement in PSNR to motion compensation in video compression. In 2009, Bovik\cite{bovik_2009} reports up to $\times 3$ improvement in bitrate for the same picture quality with more recent codecs like H.264. Our dataset here shows about 150\% improvement in bitrate for H.264 at the same PSNR cf non-motion compensated video coding, and using P-frames only.  

\subsection{Modern motion estimation and video compression}
Motion estimation since 2000 has been characterised by the quantitative application of temporal and spatial smoothness constraints. This was well articulated with the Bayesian approach by Konrad et al\cite{Konrad:1992, stiller_2001}. Other landmark work followed using optimisation strategies like Graph Cuts\cite{boykov_2001} and variational methods together with second order constraints on the motion field\cite{Bro04a}. Much of this development is tracked by the survey work of Baker et al\cite{Baker_2011}. A handful of authors have reported on the use of optic flow or modern motion estimation in codecs. 

Stiller\cite{stiller_1990} employed spatial smoothness constraints in block matching motion estimation for a codec at 8Kbits/sec. He reported about 1dB improvement at the same bitrate over standard block matching for three short sequences {\em Miss America, Swing, Alexis} which was state of the art at the time. It was not clear what codec was being used then.

Kohli et al\cite{kohli_2013} point out that rate distortion criteria can and should be integrated within the new optical flow algorithms for their use in video compression. They develop a wavelet based motion estimation algorithm using the RD criterion above, and apply it to a variant of the Dirac codec\cite{dirac}. They report gains from 1-49\% on bitrate at the same PSNR when compared to the Dirac codec with block matching. 

Taubman\cite{taubman_2015} presented a more elaborate RD scheme for motion vector estimation in a codec and deployed variational methods for optimisation. Like Kohli et al they deploy a wavelet model of the motion field. They report about 3dB improvement at the same bitrate for 4 CIF sequences {\em flower, foreman, mobile, harbour}. These gains were modeled based on measured motion compensated frame differences and not from an actual codec implementation.

Other work that employed modern optic flow approaches\cite{kameda_2016} concentrated on motion compensation using dense flow to extrapolate a new frame for prediction, rather than using the decoded frame. Gains in bitrate are reported at about 7\% when motion is steady across frames, and negative when there is unsteadiness in the shot. This makes sense since extrapolation is unlikely to be successful when temporal motion is not smooth.

\subsection{Our contributions}
This paper attempts to update the limited previous efforts to use optic flow and modern motion estimation in codecs. The work of Kroeger et al \cite{kroeger2016fast} is used as an example algorithm for testing. It appears in the SINTEL league table of motion estimators\cite{Butler:ECCV:2012} as rank 131. In comparison, Taubman's dense block matching motion estimator is at 135.   Previous work has not attempted to incorporate these algorithms within an entire working codec, and we do this with the MPEG4 and H.264 codecs implemented in FFMPEG.  We use the SINTEL\footnote{{\tt http://sintel.is.tue.mpg.de/ }} dataset\cite{Butler:ECCV:2012} with {\em ground truth motion} to validate claims about the use of true motion in video codecs.  To use flow estimation algorithms in a block based codec requires some massaging of the flow field to extract appropriate block vectors. We report on the performance of 2 techniques for doing this. A new technique for enriching the block matching search with motion information from the optic flow motion estimator is also presented. Finally, we use a wider range of material for evaluation than has been used in the past.

\section{THE DATASET}

We deploy the SINTEL sequence set\cite{Butler:ECCV:2012} for evaluation of performance with ground truth motion. This dataset was chosen as the sequences are some of the longest with available ground truth motion data\cite{Butler:ECCV:2012}. Both the albedo pass and the final pass of these scenes were used. Each of the sequences was 50 frames long with resolution 1024x436 and YUV420 pixel format.

\section{THE CODEC IMPLEMENTATIONS}
We employ {\tt ffmpeg/MP4} and {\tt libx264} for our experiments. The MPEG4 and H.264 codecs were used with the invocations as follows for MP4.
\begin{center}
\begin{minipage}{0.8\linewidth}
 {\tt
 -i <fname>.y4m -c:v mpeg4  -pix\_fmt yuv420p -bf 0  -g 100  -q 5 -subq 0 <fname>.mp4  -y -psnr -vstats
 }
\end{minipage}
\end{center}
For H.264 we use the following options
\begin{center}
\begin{minipage}{0.8\linewidth}
 {\tt CODEC\_OPTS="-x264opts no-psy=1:aq-mode=0:ref=0:subme=1:min-keyint=100"
 }
\end{minipage}
\end{center}
We employ only P frames with at most 1 P frame reference allowed, and a GOP size of 100 frames. Quantiser step sizes varied from 5 to 45 in steps of 5. 

\section{TRANSFORMING FLOW FIELDS FOR USE IN CODECS}
The output of a dense flow field motion estimation or optic flow algorithm is usually a single motion vector per pixel. The codecs we use here, H.264 and MPEG4 require a single vector per block of pixels. The blocks may vary in size from $4 \times 4$ to $16 \times 16$ in this case. The problem can be posed in a Bayesian fashion through the manipulation of $p({\bf d}_B | {\bf d}_{k\in B}, I_n, I_{n-1})$, the posterior probability of the block vector ${\bf d}_B$ given each flow vector in the block $B$, ${\bf d}_{k\in B}$ and the image material in the current and previous frames $I_n, I_{n-1}$. This can be written as 
\begin{align}
    p({\bf d}_B | {\bf d}_{k\in B}, I_B) & \propto p(I_n, I_{n-1} | {\bf d}_B) p({\bf d}_B | {\bf d}_{k\in B}) 
\end{align}
We generate the MAP estimate of ${\bf d}_B$ in a block using a candidate selection scheme where candidates arise from the flow field in the region of the block patch. Hence we evaluate the posterior probability of each flow vector in the block, and choose the most likely vector. Typically the likelihood $p(I_n, I_{n-1} | {\bf d}_B)$ would be a normal distribution of the SSE over the block given that vector and we may choose Gaussian or Gibbs priors for the vector field term. We test different estimates for ${\bf d}_B$ as follows.
\begin{center}
\begin{minipage}{0.7\linewidth}
\begin{enumerate}
    \item[Vector Median (VM)] Assuming a constant likelihood and a Gibbs smoothness prior for the motion field, the optimal estimate is the Vector Median of the $K$ flow vectors in the block. That is $\hat{\bf d}_B^V = Med({\bf d}_k : k = 1\ldots K)$.
    \item[Mean (M)] Assuming a constant likelihood and a separable Gaussian smoothness prior for the motion field, the optimal estimate is the Mean of the $K$ flow vectors in the block. That is $\hat{\bf d}_B^{M} = (1/K)\sum_{k=0}^{K-1}{\bf d}_k$.
    %\item[HMAP-V/M] Assuming a Gaussian likelihood, and either of the priors above, the optimal candidate $\hat{\bf d}_B^{ML}$ is the one minimising the sum of SSE and smoothness energy.  We reduce the candidate set to the optimal estimates VM and M above, and additionally include the result of the internal ffmpeg block matcher.  $\hat{\bf d}_B^{ML}$ is the flow vector which minimises the SSE in that block.
\end{enumerate}
\end{minipage}
\end{center}
The VM estimate is more computationally intensive than the Mean. We could have also used a MAP estimate employing both a flow field prior and the Gaussian likelihood but because the rate control in the codec would already make a decision based on image cost of some kind, we do not explore that here. It is however worth considering this in future work.

\section{USING THE FLOW VECTOR IN THE CODEC}
In the codecs we experiment with, the block matching cost function is the RD cost presented above in equation~\ref{rd}. Classically, we should also incorporate this cost in the flow estimation process. In this work we do not, hence avoiding the reworking of the optimisation of a new cost function for the complex optical flow algorithm. Furthermore, we experiment with flow vectors derived from the original sequence and not the currently decoded frame. This breaks the feedback loop of the encoder but it allows us to explore the impact of {\em true/real motion} without the artefacts of compression. 

We deploy our modification to the codec inside the motion estimation routine. When the encoder requests a motion vector to be calculated for a block, we return instead a block vector derived from our flow field, but suited to that block size according to the process outlined above. That flow field is either provided by ground truth (T0) or pre-computed on the original data (T1)  or employs the decoded previous picture to compute the flow (T2). 

In an effort to employ the motion RD cost in equation~\ref{rd} without substantially increased computational load, we experiment with evaluating two motion candidates with that cost function. One candidate is generated from the internal motion estimator ({\tt Diamond or Hex}) used in the codec  and the other is from VM, or M above. The RD cost used is the same as employed internally in the codec with $\lambda_y = 2^{q/6 - 2},~\lambda_c = q^2 * 0.9 * 256$. Where $\lambda_y, \lambda_c$ are used with luminance and colour plane data respectively. These two schemes are called {\em Hybrid Median/Mean} depending on the flow field candidate used.

\begin{figure}
    \centering
    \includegraphics[width = 0.45 \linewidth]{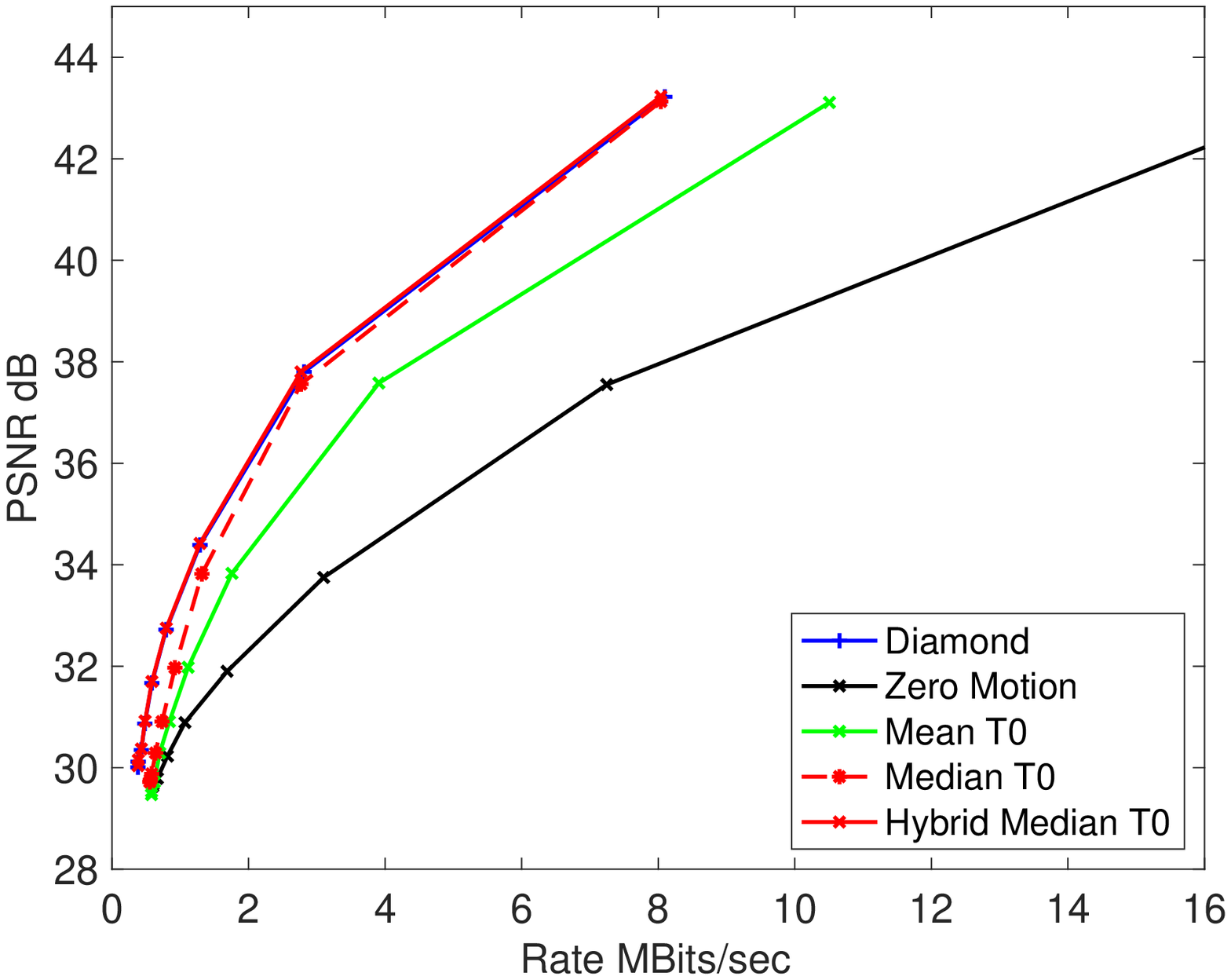} \hspace*{0.5cm}
    \includegraphics[width = 0.45 \linewidth]{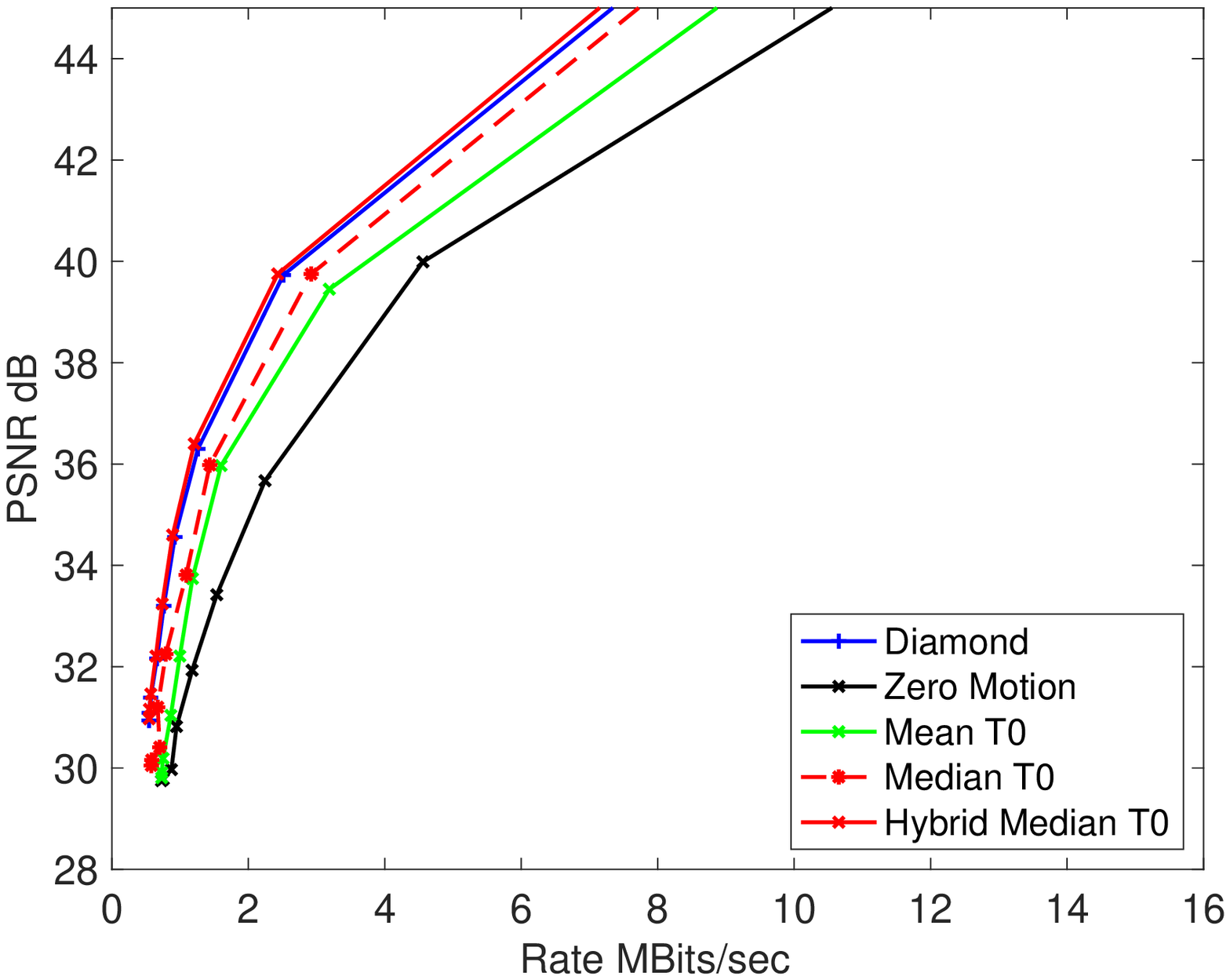}
    \caption{Aggregated RD Curves for 5 different motion estimator types compared across the Sintel Dataset using MP4. Left: Albedo render Right: Final render. The Albedo rendered scenes tend to be more cartoonish than the final renders. The final tenders contain more motion blur, lighting finishes, shadows etc. That means the performance of the codec is generally better with the final renders than the albedo renders. }
    \label{sintel-t0}
\end{figure}

\section{EXPERIMENTS WITH TRUE/GROUND TRUTH MOTION}
The SINTEL dataset contains synthetic scenes and therefore a ground truth flow field (T0) is provided for testing. Figure~\ref{sintel-t0} shows the RD curves (using the MP4 codec) aggregated over 5 Sintel sequences. We use quantiser step sizes $q = 2, 5:5:40$ and measure $r_q^i,d_q^i$ the rate and psnr at each of those quantiser settings for each sequence $i$. Each point is the median $r_q^i,d_q^i$ across all the sequences.  {\em Zero Motion} uses no motion at all, {\em Mean T0} and {\em Median T0} use the motion vector provided by the average and vector median of the motion in the block respectively, and {\em Hybrid Median T0} refers to the sub-optimal RD scheme referred to previously. {\em Diamond} uses the internal motion estimator in the MP4 codec provided by {\tt ffmpeg}. 

Is is firstly satisfying to note that using motion in a codec is much better than no motion. The BD Rate\cite{bdrate} gain is more than 150\%, and the BD PSNR gain is as much as 5dB. Surprisingly using ground truth motion alone yields a worse RD performance than {\em Diamond}.  But the block based downsampling scheme used does have an effect, with the Median vector outperforming the Mean vector. Because the motion being used is not estimated from the decoded frames, the convergence in the Median and Mean estimates at low bit rate is expected. However they always perform better than no motion at all. The Hybrid schemes perform better in general than {\em Diamond} but the improvement is small, a BD rate gain of 1.6\% and PSNR gain of 0.08 dB, for the Final renders and 1.5\%, 0.08dB for the Albedo renders. These gains vary significantly by sequence, with one sequence {\em Cave-4} showing a BDRate gain of 4.4\% and a BDPSNR gain of 0.2dB, while {\em Alley-1} has a rate gain of just 0.7\% and PSNR gain of 0.04dB.

The fact that the encoder is being tested with motion in an open loop is probably affecting the observations, but it is very surprising that over 40dB the {\em Diamond} scheme still performs well wrt massaging the ground truth motion field. 

\begin{figure}
    \centering
    \includegraphics[width = 0.45 \linewidth]{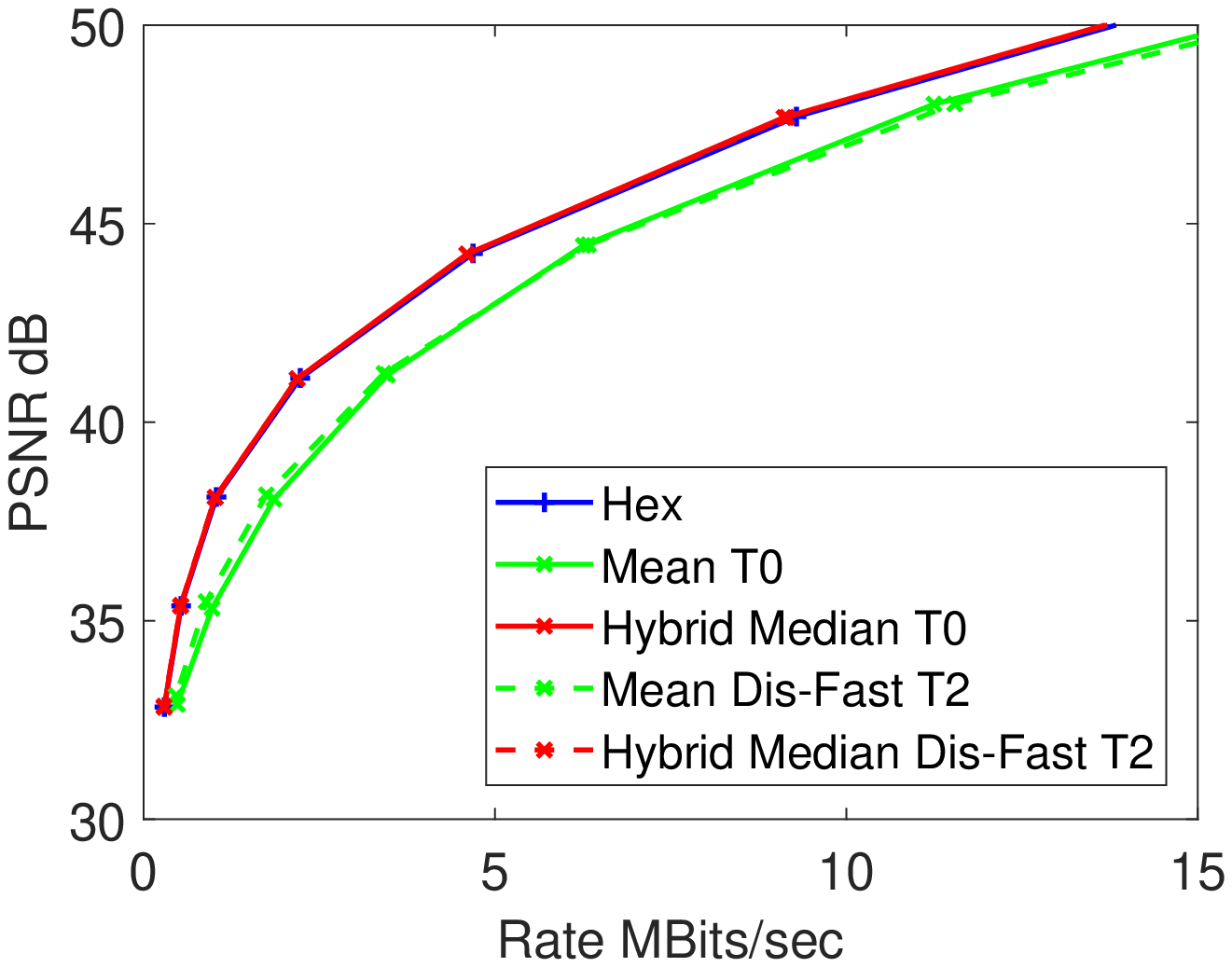} \hspace*{0.5cm}
    \includegraphics[width = 0.45 \linewidth]{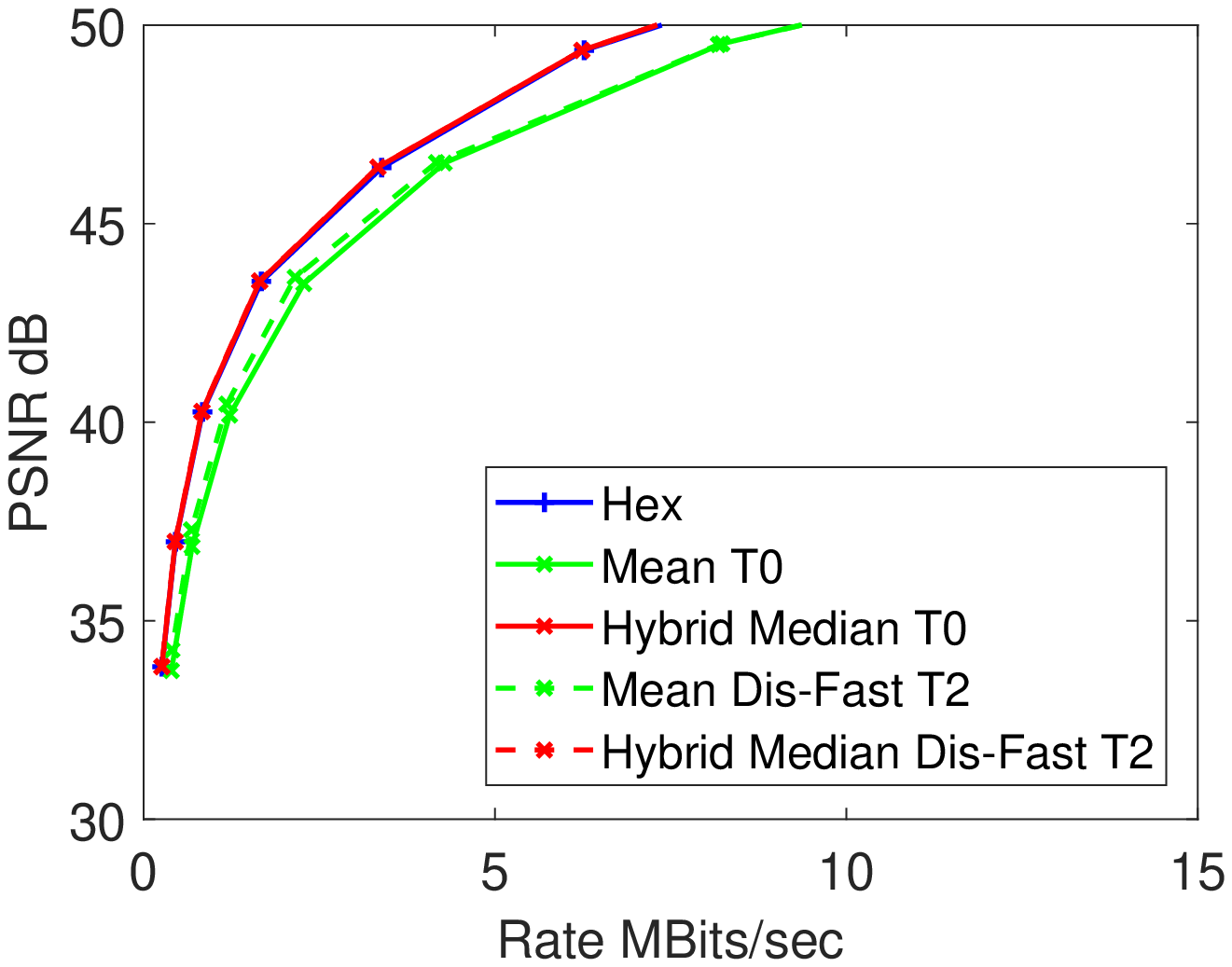}
    \caption{Aggregated RD Curves for 3 different motion estimator types compared across the Sintel Dataset using x264. Left: Albedo render Right: Final render. The performance of the codec is generally better with the final renders than the albedo renders. }
    \label{x264}
\end{figure}

\section{EXPERIMENTS WITH REAL MOTION IN A CLOSED LOOP ENCODER}
 For our experiments with a recent optic flow estimator, we employ the optic flow  estimators of Kroeger et al\cite{kroeger2016fast} (Dis-Fast) and Weinzapfael et al\cite{deepFlow2013} (DeepFlow). These rank 133 and 63 respectively (as of July 2018) in the Sintel optic flow league table. The only other Block Matching algorithm compared, Xu and Taubman\cite{xu_2013} ranks 137. We chose Dis-Fast because of its speed, it is the fastest in that list; while DeepFlow is one of the better known recent estimators. The Diamond or Hex motion estimators used by MP4 and H.264 in {\tt ffmpeg/x264} have not yet been compared on that list. In this set of experiments we use Dis-Fast and DeepFlow in closed loop, on the previously decoded frames i.e. in mode T2.

Figure~\ref{x264} shows results over the same set of files as used previously. We present a limited set of motion types here {\em Hex, Mean-T0, Hybrid-Median, Dis-Fast} to focus on the key points. The same overall observation persists, that using the ground truth motion field {\em Mean T0} yields worse results overall than motion estimation using the decoded frames T2. However, using Dis-Fast in a closed loop yields better RD performance than the ground truth motion. The gains with the Hybrid schemes using Dis-Fast in closed loop are more in H.264 than in MP4. Also of interest is that the vector median outperforms the mean. The BD Rate gain of {\em Hybrid Median Dis-Fast} is 1.53\% on average across the sequences for the Final renders and 1.19\% for the Albedo renders. The BD PSNR gain is 0.08 dB in both cases. This performance varies quite a bit across sequences. The table below shows these metrics for different algorithms and different sequences compared to the performance with {\tt Hex} in {\tt ffmpeg/x264} itself. In that table {\em Dis} refers to Dis-Fast and {\em Hy} is used to indicate the Hybrid algorithm. The BD PSNR gains for the various algorithms are about the same at 0.09 dB for both T0 and T2.
\begin{center}
     BD Rate \% gains versus {\tt Hex} for Final Renders in H.264 {\small
    \begin{tabular}{|r|c|c|c|c|c|c|c|c|c|}\hline
         Motion & Alley 1 & Ambush 5 & Bamboo 2 & Sleeping 2 &  Mountain 1 & Shaman 3 & Temple 2 & Average \\ \hline
         Mean T0 & -18.79 &   -5.48  &  -15.77 & -18.28  & -13.92 & -18.52 & -16.64 & -12.05 \\
         Median T0 & -16.17 & -4.31 &  -11.14 & -14.12  &  -11.4 &  -16.59 &  -11.57 & -10.1 \\
         Hy-Mean T0 & 2.36  &  0.36 &  0.73 &  1.51  &  1.85 &  2.16 &    1.46 & 1.49 \\ 
         {\bf Hy-Med T0} & 2.4 & 0.44 & 0.89 &  1.55  &  2.04 & 2.24 & 1.63 & {\bf 1.6} \\
         Hy-Mean Dis T2 & 2.31 &   0.5 &    0.8 &    1.63  &    1.74 &    1.87 &   1.39 & 1.46 \\
         {\bf Hy-Med Dis T2} & 2.41 &  0.53 &  0.85 &   1.74  & 1.91 &    1.84 & 1.45 & {\bf 1.53} \\ \
         {Hy-Mean DF T2} & 2.460 & 0.48 & 0.94 & 1.84 & 1.89  & 2.15 &   1.58 & 1.62 \\
         \bf{Hy-Med DF T2} & 2.54 & 0.5 & 1.02 &  1.83 &  1.83  &  2.05 &   1.70 & {\bf 1.63} \\ \hline
    \end{tabular} }
\end{center}
The last two rows also show the Hybrid algorithm performance using the DeepFlow (DF) motion estimator. The BD Rate gain here is on average 1.63\% which is the best overall. The corresponding BD PSNR gain is 0.1 dB. Note that for some sequences the rate gains are about 2\%. 

%The situation is more encouraging if we analysis low bitrates separately from high bitrates. At low bitrates, below 1Mbits/sec the gains with better motion estimation in H.264 become more significant e.g. for {\em Hybrid Mean Dis-Fast} the BD PSNR gain is 1.4dB below 1 Mbps and just 1.4\% above 1Mbps averaged across the sequences. The BD PSNR gain is steady at 0.02 dB across those intervals.

\section{DISCUSSION AND FINAL COMMENTS}

In general, these results encourage further thought. When used inside a codec with rate control, gains with the use of an optic flow estimator seem to disappear. Rate gains seem to be about 1-4\% depending on the sequence and the codec, while PSNR gains are modest at about 0.1dB. In many ways, the results do not reflect the promise of previous work which report gains of the order of 1dB or as much as 49\% in rate.

\begin{figure}
    \centering
    \includegraphics[width = 0.45 \linewidth]{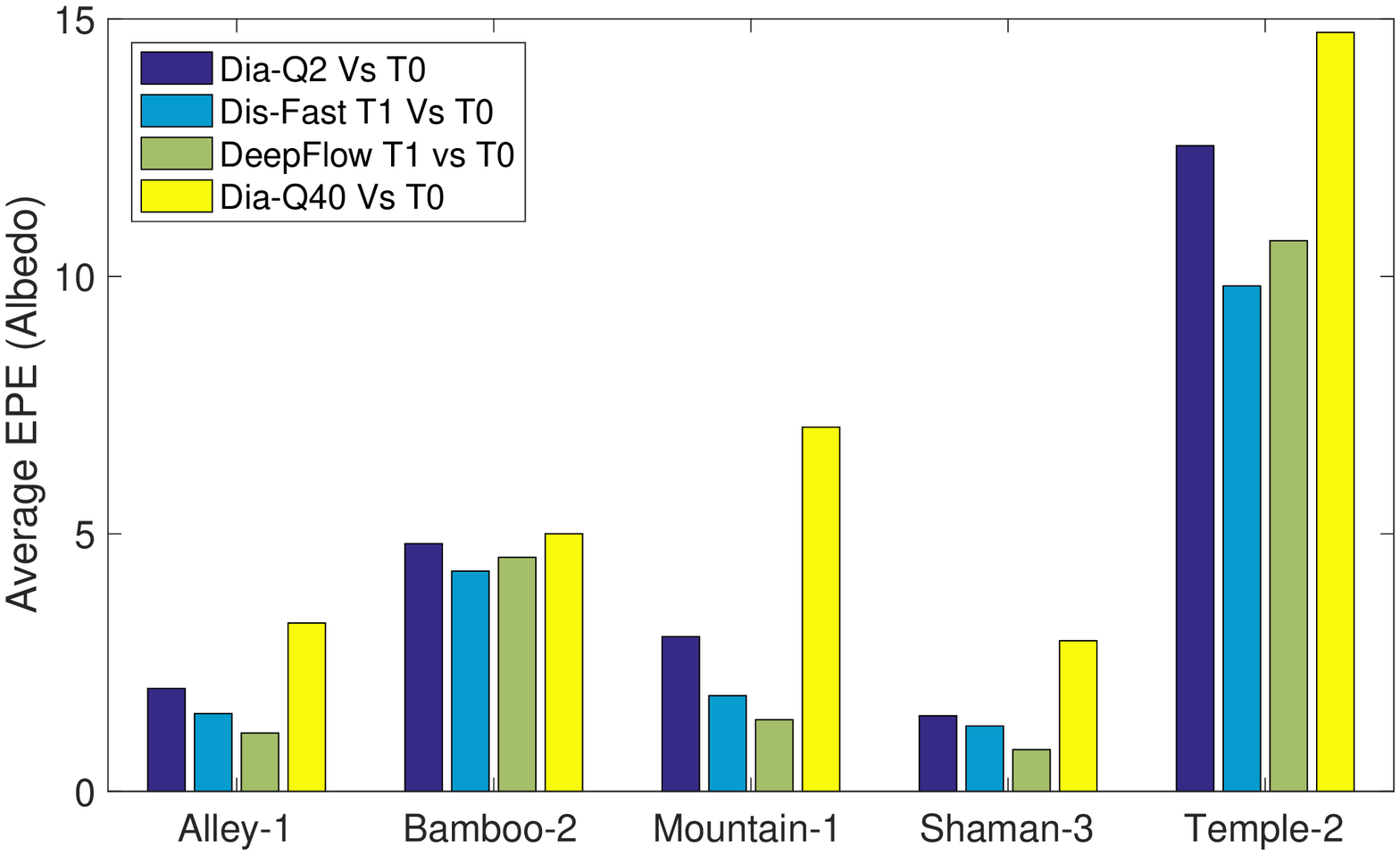} \hspace*{0.5cm}
    \includegraphics[width = 0.45 \linewidth]{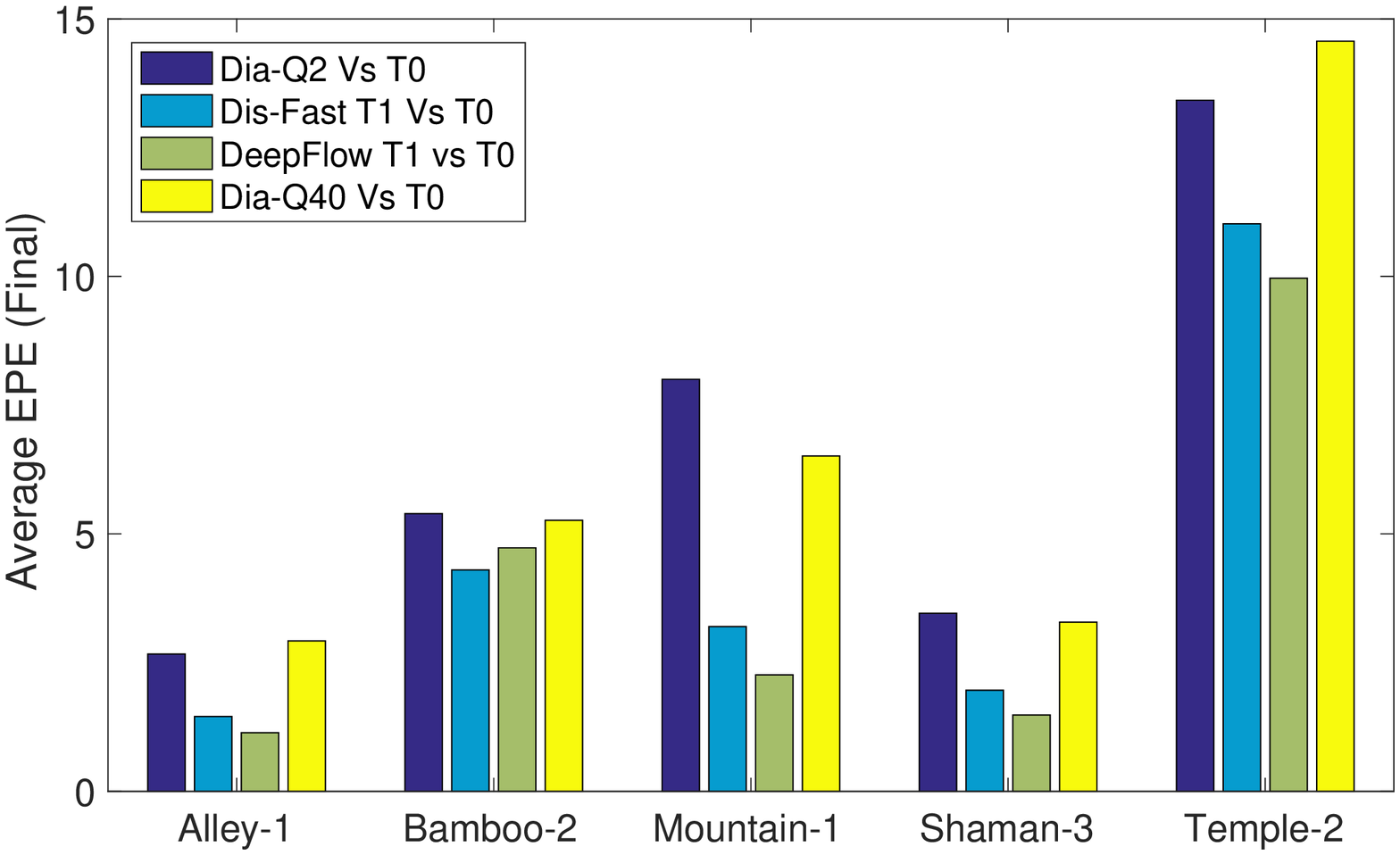} \\
        \includegraphics[width = 0.45 \linewidth]{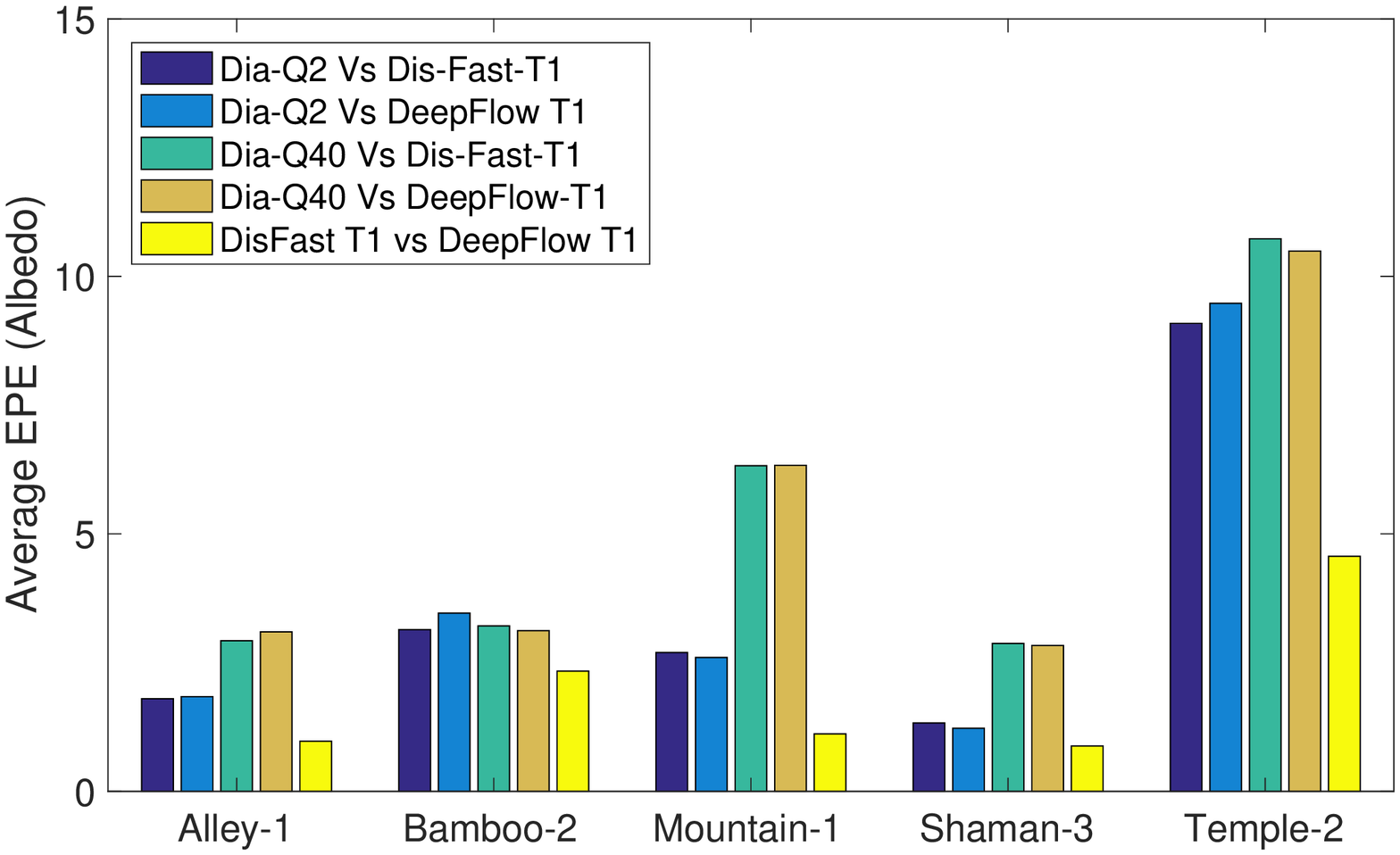} \hspace*{0.5cm}
    \includegraphics[width = 0.45 \linewidth]{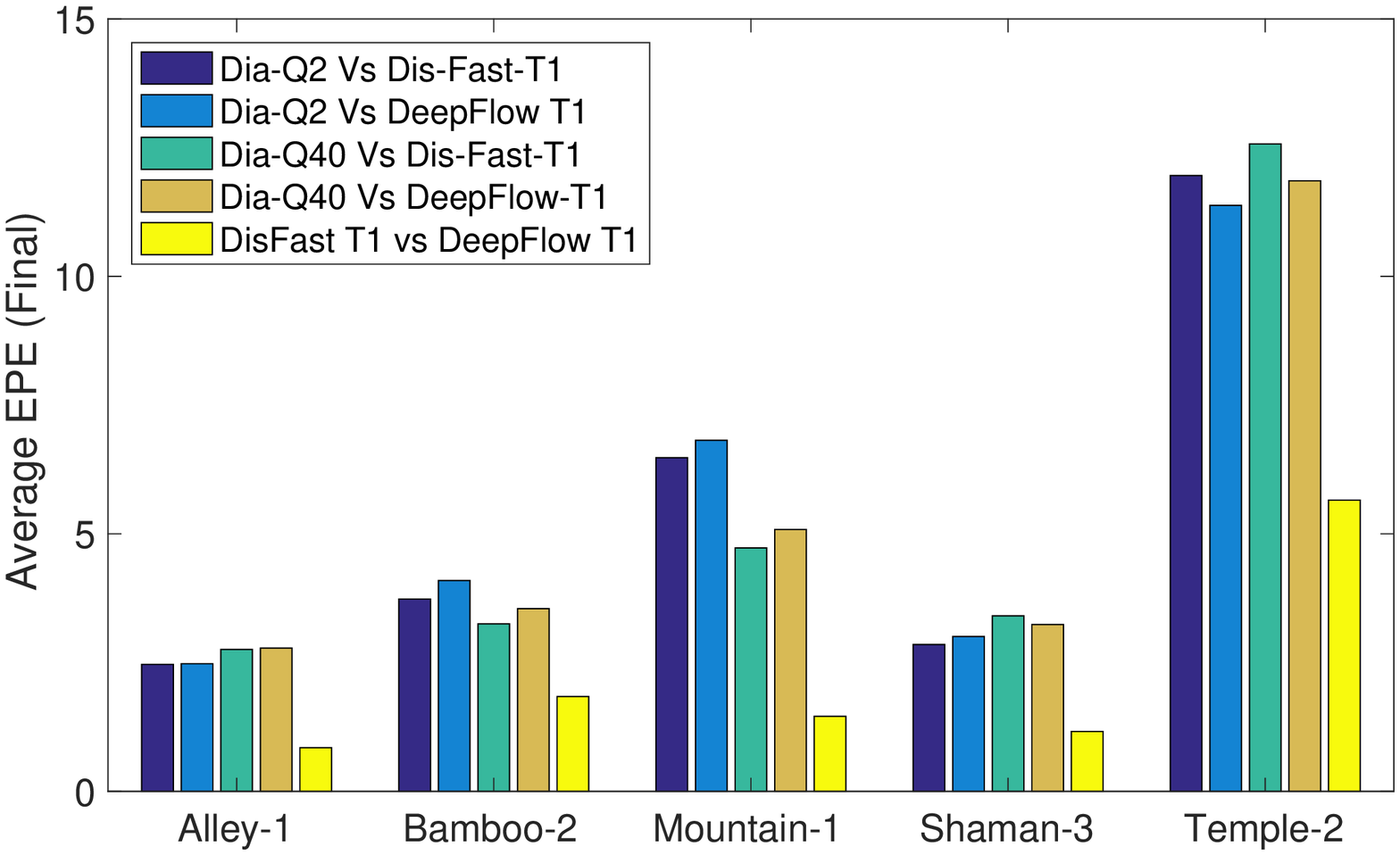} 
    \caption{Showing End Point Error for 5 Sintel sequences using 3 different motion types compared to ground truth (T0) and each other. Top Row : differences with ground truth, Bottom Row : differences between each other. DeepFlow yields the closest result to ground truth. At a quantiser value of 40 Diamond deviates from T0 much more than at $q=2$. That is sensible because $q=40$ implies a much lower picture quality and hence the estimated motion vectors would not necessarily be well matched to ground truth. The differences in the bottom row are an indicator that there is diversity in the motion estimates. }
    \label{epe}
\end{figure}

Figure~\ref{epe} shows a comparison between the motion fields estimated by the  {\em Diamond} search block matcher in ffmpeg, Dis-Fast (T1), DeepFlow (T1) and ground truth (T0). We see that DeepFlow is indeed best as far as ground truth is concerned, with Dis-Fast second best. In some sequences (Bamboo and Temple) though, they are about as different from ground truth as Diamond. The bottom row also shows that for those sequences they are also different from each other. Those sequences show complicated/fast motion and textures scenes. This might imply that Dis-Fast and Deep Flow are just not that different enough to generate much better performance than {\em Diamond} inside a codec for those scenes. Those are the scenes with very small BD gains.

The optic flow algorithm here was incorporated into the codec in a rather naive fashion, ignoring RD motion criterion. The fact that the Hybrid estimators outperformed Diamond (albeit by a small margin) shows that using RD criterion in the optic flow optimiser may be the best approach to increase the gains. 

Finally, this work considered P frames only, and in a long GOP size (100 frames). That means that decoded frames can expect to deviate significantly from the I frame by the end of the GOP. Because we use only P-frames, that means that the direct search used in the codec may have a better chance of reducing prediction error than the implicit search with smoothness constraints of the optic flow motion estimator. 

We shall continue to develop this experimental work in the future. It is sensible to consider next, motion estimators higher up in the rank than Dis-Fast or DeepFlow, a more useful parameter set for these algorithms (they were deployed out of the box here); and the impact of the use of B-frames for resolving occlusion/uncovering. 

\acknowledgments % equivalent to \section*{ACKNOWLEDGMENTS}       
 
This work was supported in part by YouTube, Google and the Ussher Research Studentship from Trinity College.

% References
\bibliography{motion} % bibliography data in report.bib
\bibliographystyle{spiebib} % makes bibtex use spiebib.bst

\end{document}